\documentclass{article}
\usepackage[dvips]{graphics}
\newcommand{\Rset}{{\mathbb R}}

\newcommand{\diag}{{\rm diag}}

\newcommand{\beq}{\begin{equation}}
\newcommand{\bea}{\begin{eqnarray}}
\newcommand{\ba}{\begin{array}}
\newcommand{\eeq}{\end{equation}}
\newcommand{\eea}{\end{eqnarray}}
\newcommand{\ea}{\end{array}}
\newcommand{\bean}{\begin{eqnarray*}}
\newcommand{\eean}{\end{eqnarray*}}

\newcommand{\ft}[2]{{\textstyle {\frac{#1}{#2}} }}

\newcommand{\sO}{{\mathcal O}}

\renewcommand\thesection{\Roman{section}}

\usepackage{amsfonts, latexsym, amssymb}

\begin{document}

\title{Fourier transforms of Lorentz invariant functions}
\author{Alexander Wurm\\
 {\small Center for Relativity and Department of Physics}\\
 {\small  University of Texas, Austin, TX 78712, USA}\\[0.25in] 
Nurit Krausz\thanks{Now at: Schonfeld Securities, 650 Madison Ave, New York,
 NY 10022.} \\ {\small Department of Mathematics}\\
 {\small University of Texas, Austin, TX 78712, USA}\\[0.25in]
C\'ecile DeWitt-Morette and Marcus Berg \\
 {\small Center for Relativity and Department of Physics}\\
 {\small University of Texas, Austin, TX 78712, USA}}

\date{}

\maketitle

\begin{abstract}
Fourier transforms of Lorentz invariant functions in Minkowski space, with
support on both the timelike and the 
spacelike domains are performed by means of direct integration.
The cases of $1+1$ and $1+2$ dimensions are worked out in detail,
and the results for $1+n$ dimensions are given.
\end{abstract}

\pagebreak

\section{Introduction}
The main goal of this paper is to perform spherical averages over
the hyperbolic coordinates in Minkowski spacetime by means of direct
integration. We work with Lorentz invariant (or ``radial'') functions,
defined on ${\mathbb R^{1,n}}$,
which depend only on the distance to the light-cone $s^2 = \left(x^0
\right)^2 -\left( \vec{x}\right)^2$. ($x^0$ is the time coordinate
and $\vec{x}$ stands for the spatial coordinates).
In particular, we will
 compute their Fourier transforms:
\beq
F(k) := \int_{{\mathbb R^{1,n}}} dx^{n+1}\; f\!\left(\eta_{\mu \nu} x^{\mu}
x^{\nu}\right)
\; \exp\!\left( - 2\pi i k_{\mu} x^{\mu}\right), \label{fourier} 
\eeq
where the metric $\eta_{\mu\nu}$ is of the form
\beq
\eta_{\mu \nu} = \diag\left( 1, -1,\ldots ,-1\right)\label{metric}
\eeq

(We set the $2\pi$ in the exponential to avoid factors of $2\pi$
appearing in front of the integral. Here it is just for convenience but
this strategy becomes crucial when
integrating over infinite-dimensional spaces.${}^1$)

This type of integral is ubiquitous in Quantum Field Theory (QFT),
 since the fields can be decomposed in terms of their Fourier components,
and the correlation functions depend only on the radial distance $s^2$.
Although QFT is defined in Minkowski space, a common procedure is
to perform a Wick rotation at an early stage (this procedure can be
found in standard textbooks, e.g., Ref.~2) in which all integrals
 over spacetime are reduced to Euclidean integrals.

There are several obvious disadvantages to relying on the Wick rotation
procedure, in which the time coordinate $t$ is rotated to imaginary
values $t\to -it$, and Minkowski spacetime is transformed into Euclidean
space. First, the special structure of Minkowski spacetime in which
there is a preferred direction (i.e., time direction) is lost. Second, Wick
 rotation might not be a valid procedure in all circumstances, e.g., in
 cases where the arc at infinity does not vanish.

The motivation for this work is to provide tools for performing 
calculations in QFT while working directly in Minkowski space.

In analogy with Euclidean space, we introduce a (pseudo-) spherical
coordinate atlas, i.e., one of the angles is hyperbolic.
Integrating over all the angular variables leaves
us with a one-dimensional radial integral. The advantage of this procedure
is that any possible light-cone singularities of the integrand are mapped to
point singularities in the radial variable and therefore easier to deal
with.

In this paper we will not specify the function space for which the
integrals are convergent. We will assume that only such functions have been
chosen.

Below, we will discuss the cases $n=1$ and $n=2$ in detail. For the
case of general $n$ we only quote the results.

 After preparing an earlier
version of this paper we found an article by Codelupi${}^3$ which
seems to be largely unknown. Codelupi studied Fourier transforms of
Lorentz invariant functions $f(s)$, using related methods, and found
the same results. To find the case of arbitrary $n$ he proved a recursion
relation, that relates Fourier transform of $n+2$ spatial dimensions to the
Fourier transform of $n$ spatial dimensions. In Section IV we adapt
 Codelupi's elegant method to our case.

\section{Case $\Rset^{1,1}$ }

\subsection{Pseudospherical coordinate atlas for $\Rset^{1,1}$}

A characteristic feature of manifolds with an indefinite metric is that
a global spherical coordinate system does not exist.${}^4$ To
cover $\Rset^{1,1}$ we use four patches and four different parametrizations.
The parametrizations coincide on the boundary between the domains, i.e.,
on the light-cone. (See Fig.~1).
\begin{figure}[h]
  \begin{center}
    \resizebox{4cm}{!}{\includegraphics{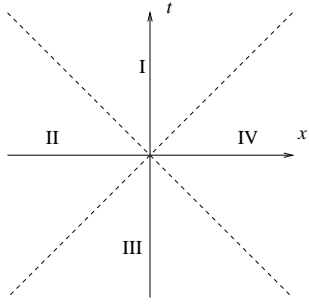}}
    \caption{Standard division of $\Rset^{1,1}$.
     I is the forward light-cone, III the
     backward light-cone, and II and IV the spacelike domains.}
   \end{center}
   \label{fig:label}
\end{figure}

Denote by  $t,x$ the global Cartesian coordinate system with
 distance $s^2=t^2-x^2$. We parametrize the different patches as follows:

\begin{itemize}
\item Patch I: 
\[
t = s \cosh \psi,  \qquad\qquad\qquad x=s \sinh \psi,
\]
where $s\in [0,+\infty[$ and $\psi\in ]-\infty,+\infty[$.\\
Volume element:  $dt\wedge dx = s\, ds\wedge d\psi$.\\
Line element: $s^2 = t^2-x^2$. 

\item Patch II:
\[
t=i s \sinh \psi, \qquad\qquad\qquad x=i s \cosh \psi, 
\]
where $s\in ]i\infty, i0]$ and  $\psi\in ]-\infty,+\infty[$.\\
Volume element:  $dt\wedge dx = s\, ds\wedge d\psi$.\\
Line element: $s^2= t^2-x^2$. 

\item Patch III:
\[
t = s \cosh \psi,  \qquad\qquad\qquad x=s \sinh \psi,
\]
where $s\in [0,-\infty[$ and $\psi\in ]-\infty,+\infty[$.\\
Volume element:  $dt\wedge dx = s\, ds\wedge d\psi$.\\
Line element: $s^2= t^2-x^2$. 

\item Patch IV:
\[
t=i s \sinh \psi, \qquad\qquad\qquad x=i s \cosh \psi, 
\]
where $s\in ]-i\infty, i0]$ and  $\psi\in ]-\infty,+\infty[$.\\
Volume element:  $dt\wedge dx = s\, ds\wedge d\psi$.\\
Line element: $s^2= t^2-x^2$. 

\end{itemize}
The limits of integration in each patch are chosen to yield a positive
result when the volume element is integrated over a small, finite volume.

\subsection{Fourier transform of radial functions}

We integrate the function $f\!\left(s^2\right)$ separately in the timelike and spacelike
domain for the cases of timelike and spacelike momenta. To simplify the
calculation, we set $k_x=0$ when the momentum is timelike, and $k_t=0$
when the momentum is spacelike. This can always be achieved with a Lorentz
transformation and is not a restriction on the results.

\begin{itemize}

\item Patch I+III:
\begin{itemize}
\item[(i)] Timelike momentum: $k_x=0$,
\bea
I_{I+III}(k_t) & = & \int_{I+III} dt\, dx\; f\!\left(t^2-x^2\right)\;
\exp\!\left( -2\pi i k_t t\right)\nonumber\\
& = & \int_0^{\infty} ds\; s\; f\!\left(s^2\right) \int_{-\infty}^{+\infty}
d\psi\; \exp\!\left(-2\pi i k_t s \cosh \psi\right)\nonumber\\
& & + \int_0^{-\infty} ds\;
s\; f\!\left(s^2\right) \int_{-\infty}^{+\infty} d\psi\;\exp\!\left(-2\pi i k_t s \cosh \psi
\right)\nonumber\\
& = & 2 \int_0^{\infty}ds_0\; s_0\; f\!\left(s_0^2\right) \int_{-\infty}^{+\infty} d\psi\;
 \cos\!\left( 2\pi k_t s_0\cosh\psi\right)\nonumber\\ 
& = & - 2\pi \int_0^{\infty} ds_0\; s_0\; f\!\left(s_0^2\right)\; 
N_0 \left(2\pi k_t s_0\right),
\eea
where $N_0$ is a Bessel function of zeroth order and
\beq
s_0 = \sqrt{t^2-x^2}.
\eeq
In the last line we have used formula 3.868(2) from Ref.~5, after a change
of variable to $x=e^{\psi}$.
\item[(ii)] Spacelike momentum: $k_t=0$,
\bea
I_{I+III}(k_x) & = & \int_{I+III} dt\, dx\; f\!\left(t^2-x^2\right)\;
\exp\!\left( -2\pi i k_x x\right)\nonumber\\
& = & \int_0^{\infty} ds\; s\; f\!\left(s^2\right) \int_{-\infty}^{+\infty}
d\psi\; \exp\!\left(-2\pi i k_x s \sinh \psi\right)\nonumber\\
& & + \int_0^{-\infty} ds\;
s\; f\!\left(s^2\right) \int_{-\infty}^{+\infty} d\psi\;\exp\!\left(-2\pi i k_x s \sinh \psi
\right)\nonumber\\
& = &  \int_0^{\infty}ds_0\; s_0\; f\!\left(s_0^2\right) \int_{-\infty}^{+\infty} d\psi\;
 \cos\!\left( 2\pi k_x s_0\sinh\psi\right)\nonumber\\ 
& = & 4 \int_0^{\infty} ds_0\; s_0\; f\!\left(s_0^2\right)\; K_0 \left(2\pi k_x s_0\right),
\eea
where $K_0$ is a Bessel function of zeroth order and $s_0$ as before.
In the last line we have used formula 3.868(4) from Ref.~5 , after a change
of variable to $x=e^{\psi}$.
\end{itemize}
\item Patch II+IV:
\begin{itemize}
\item[(i)] Timelike momentum: $k_x=0$,
\bea
I_{II+IV}(k_t) & = & \int_{II+IV} dt\, dx\; f\!\left(t^2-x^2\right)\;
\exp\!\left( -2\pi i k_t t\right)\nonumber\\
& = & \int_{i\infty}^{i0} ds\; s\; f\!\left(s^2\right) \int_{-\infty}^{+\infty}
d\psi\; \exp\!\left(2\pi  k_t s \sinh \psi\right)\nonumber\\
& & + \int_{-i\infty}^{i0} ds\;
s\; f\!\left(s^2\right) \int_{-\infty}^{+\infty} d\psi\;\exp\!\left(2\pi k_t s \sinh \psi
\right)\nonumber\\
& = &  \int_0^{\infty}ds_1\; s_1\; f\!\left(s_1^2\right) 
\int_{-\infty}^{+\infty} d\psi\;
 \cos\!\left( 2\pi k_t s_1\sinh\psi\right)\nonumber\\ 
& = & 4 \int_0^{\infty} ds_1\; s_1\; f\!\left(s_1^2\right)\;
 K_0 \left(2\pi k_t s_1\right),
\eea
where $K_0$ is a Bessel function of zeroth order and
\beq
s_1 = \sqrt{x^2-t^2}.
\eeq
In the last line we have used formula 3.868(4) from Ref.~5, after a change
of variable $x=e^{\psi}$.
\item[(ii)] Spacelike momentum: $k_t=0$,
\bea
I_{II+IV}(k_x) & = & \int_{II+IV} dt\, dx\; f\!\left(t^2-x^2\right)\;
\exp\!\left( -2\pi i  k_x x\right)\nonumber\\
& = & \int_{i\infty}^{i0} ds\; s\; f\!\left(s^2\right) \int_{-\infty}^{+\infty}
d\psi\; \exp\!\left(2\pi  k_x s \cosh \psi\right)\nonumber\\
& & + \int_{-i\infty}^{i0} ds\;
s\; f\!\left(s^2\right) \int_{-\infty}^{+\infty} d\psi\;
\exp\!\left(2\pi k_x s \cosh \psi
\right)\nonumber\\
& = & 2 \int_0^{\infty}ds_1\; s_1\; f\!\left(s_1^2\right) 
\int_{-\infty}^{+\infty} d\psi\;
 \cos\!\left( 2\pi k_x s_1\cosh\psi\right)\nonumber\\ 
& = & - 2\pi \int_0^{\infty} ds_1\; s_1\; f\!\left(s_1^2\right)
\; N_0 \left(2\pi k_x s_1\right),
\eea
where $N_0$ is a Bessel function of zeroth order and $s_1$ as before.
In the last line we have used formula 3.868(2) from Ref.~5, after a change
of variable to $x=e^{\psi}$.
\end{itemize}
\end{itemize}
In summary\\

\begin{center}
\fbox{\parbox{13cm}{\bean
I\left(k_t\right) & = & -2\pi\int_0^{\infty} ds_0\;
 s_0\; f\!\left(s_0^2
\right)\; N_0\left( 2\pi k_t s_0\right) + 4 \int_0^{\infty} ds_1\; s_1\;
 f\!\left(s_1^2\right)\; K_0 \left(2\pi k_t s_1\right)\\
I\left(k_x\right) & = & 4 \int_0^{\infty} ds_0\; s_0\; f\!\left(s_0^2\right)\;
K_0\left( 2\pi k_x s_0\right) - 2\pi \int_0^{\infty} ds_1\; s_1\; f\!\left(
s_1^2\right)\; N_0\left( 2\pi k_x s_1\right)
\eean}}
\end{center}

\subsection{Example: Fourier transform of a Gaussian}

In this section we apply the results from above to a specific
test function and show that the correct answer is obtained. 
Whenever necessary,
we define the integral $\int_0^{\infty} dx f(x)$ 
as   $\lim_{\epsilon \rightarrow 0} \int_0^{\infty} dx \, e^{-\epsilon
x^2} f(x)$.
With this proviso, we can directly compute the Fourier 
transform of a ``Gaussian'' (where the quotation marks remind us that
the exponential is imaginary): 
\[
\int dt \, dx \, e^{i(t^2-x^2)}e^{-2\pi i k_t t} = \pi e^{-i \pi^2 k_t^2}.
\]
We can now check that our Fourier transform integrals give the same result,
\bean
&& I_1\left( k_t\right) = 4\int_0^{\infty} dr
\, r e^{-(\epsilon+i)r^2} K_{\nu}(2\pi r k_t)  \\
&& \hspace{3cm} ={1 \over \pi k_t } {1 \over \sqrt{\epsilon+i}}
\, \Gamma(1+\ft{\nu}2) \Gamma(1-\ft{\nu}2)
e^{{\pi^2 k_t^2 \over 2(\epsilon + i)}} \, 
W_{\!-\ft12,\ft{\nu}2}\!\left({\pi^2 k_t^2 \over 2(\epsilon + i)}\right),
\eean
\bean
&& I_2\left(k_t\right) = -2\pi \int_0^{\infty} dr
\, r e^{-(\epsilon-i)r^2} N_{\nu}(2\pi r k_t)  \\
&& \hspace{5mm} =- {1 \over k_t } {1 \over \sqrt{\epsilon-i}}
\, { 1\over \sin({\pi \nu \over 2})} 
e^{-{\pi^2 k_t^2 \over 2(\epsilon -
 i)}}
\left[ W_{\!\ft12,\ft{\nu}2}\!\left({\pi^2 k_t^2 \over 2(\epsilon -
i)}\right)
-\cos({\ft{\nu \pi}2})\, {\Gamma(1+\ft{\nu}2) \over \Gamma(1+\nu)}\, 
M_{\!\ft12,\ft{\nu}2}\!\left({\pi^2 k_t^2 \over 2(\epsilon - i)}\right)
\right],
\eean
where we have used 6.631(2,3) from Ref.~5.
Each integral has a pole as $\nu \rightarrow 0$;
the $W$ function with negative first argument
has a simple pole in $\nu$, and there is an inverse sine in $\nu$
in the second integral. However, as we will see, this singularity
exactly cancels between the two integrals. Thus we may, in fact, take
$\nu \rightarrow 0$ in the sum.
First we convert the $M$ function to $W$ functions,
\[
M_{\ft12,\ft{\nu}2}(z) = i{\Gamma(1+\nu) \over \Gamma(\ft{\nu}2)}
W_{-\ft12,\ft{\nu}2}(z) + {\Gamma(1+\nu) \over \Gamma(1+\ft{\nu}2)}
e^{-i\pi \nu/2} W_{\ft12,\ft{\nu}2}(z)
\]
using formula 9.233(1) in Ref.~5.
Then we make use of the following identities (9.234(1,2) and 9.235 in
Ref.~5):
\bean
W_{\ft12,\ft{\nu}2}(z)&=&{\sqrt{z} \over 2}\, 
W_{0,\ft{1+\nu}2}(z) + {\sqrt{z} \over 2}\, W_{0,\ft{1-\nu}2}(z), \\
W_{-\ft12,\ft{\nu}2}(z)&=&{2\sqrt{z} \over \nu}W_{0,\ft{1+\nu}2}(z), \\
W_{0,\ft12}(z) &=& e^{-z/2}.
\eean
Expanding the $\Gamma$ functions and $1/\sin z = 1/z + z/6 +
\sO(z^2)$,
we can explicitly verify that the singularity cancels:
\[
I_1 + I_2 = -{2i \over \nu} + {2i \over \nu} +\sO(1)
= 0 +\sO(1)
\]
We have verified numerically that the constant term is, in fact,
\[
I_1 + I_2 = \pi e^{-i \pi^2 k^2}.
\]

\section{Case $\Rset^{1,2}$ }

\subsection{Pseudospherical coordinate atlas for  $\Rset^{1,2}$}

Global coordinate system: $t,x,y$, distance $s^2=t^2-x^2-y^2$.

\begin{itemize}
\item Patch I: 
\[
t = s \cosh \psi,  \qquad\qquad x=s \sinh \psi \cos\theta,\qquad\qquad
y=s \sinh \psi \sin\theta,
\]
where $s\in [0,+\infty[$, $\psi\in ]-\infty,+\infty[$ and $\theta\in [-\pi/2,
\pi/2 ]$.\\
To avoid problems when $\sinh \psi$ switches sign at $0$, the
integral over $\psi$ needs to be split up into two integrals:
$\psi\in[0,+\infty]$ and $\psi\in[0,-\infty]$.\\
Volume element:  $dt\wedge dx\wedge dy = s^2\, \sinh \psi\, ds\wedge d\psi
\wedge d\theta$.\\
Line element: $s^2 = t^2-x^2-y^2$.

\item Patch II:
\[
t=i s \sinh \psi, \qquad\qquad x=i s \cosh \psi \cos \theta,\qquad\qquad
y= i s \cosh \psi \sin\theta, 
\]
where $s\in [i0,i\infty[$, $\psi\in ]-\infty,+\infty[$ and $\theta\in
[-\pi/2 , \pi/2 ]$.\\
Volume element:  $dt\wedge dx\wedge dy = i s^2\, \cosh \psi\, ds\wedge d\psi
\wedge d\theta$.\\
Line element: $s^2= t^2-x^2-y^2$. 

\item Patch III:
\[
t = s \cosh \psi,  \qquad\qquad x=s \sinh \psi \cos\theta,\qquad\qquad
y=s \sinh \psi \sin\theta,
\]
where $s\in [0,-\infty[$, $\psi\in ]-\infty,+\infty[$ and $\theta\in [-\pi/2,
\pi/2 ]$.\\
To avoid problems when $\sinh \psi$ switches sign at $0$, the
integral over $\psi$ needs to be split up into two integrals:
$\psi\in]+\infty,0]$ and $\psi\in]-\infty,0]$.\\
Volume element:  $dt\wedge dx\wedge dy = s^2\, \sinh \psi\, ds\wedge d\psi
\wedge d\theta$.\\
Line element: $s^2 = t^2-x^2-y^2$.

\item Patch IV:
\[
t=i s \sinh \psi, \qquad\qquad x=i s \cosh \psi \cos \theta,\qquad\qquad
y= i s \cosh \psi \sin\theta,
\]
where $s\in ]-i\infty, i0]$, $\psi\in ]-\infty,+\infty[$ and $\theta\in
[-\pi/2 , \pi/2 ]$.\\
Volume element:  $dt\wedge dx\wedge dy = i s^2\, \cosh \psi\, ds\wedge d\psi
\wedge d\theta$.\\
Line element: $s^2= t^2-x^2-y^2$.
\end{itemize}

The limits of integration in each patch are chosen to yield a positive
result when the volume element is integrated over a small, finite volume.

\subsection{Fourier transform of radial functions}

As before, we integrate the function $f\!\left(s^2\right)$ separately in the
 timelike
 and spacelike domain for the cases of timelike and spacelike momenta.
 To simplify the
calculation, we set $k_x=k_y=0$ when the momentum is timelike, and 
$k_t=k_y=0$ when the momentum is spacelike. 
This can always be achieved with a Lorentz transformation and is not a
 restriction on the results.

\begin{itemize}

\item Patch I+III:
\begin{itemize}
\item[(i)] Timelike momentum: $k_x=k_y=0$,
\bea
I_{I+III}(k_t) & = & \int_{I+III} dt\, dx\, dy\; f\!\left(t^2-x^2-y^2\right)\;
\exp\!\left( -2\pi i k_t t \right)\nonumber\\
& = & \pi \int_0^{\infty} ds\; s^2\; f\!\left(s^2\right) \bigg[\int_0^{\infty}
d\psi\; \sinh\psi \exp\!\left(-2\pi i k_t s \cosh \psi\right)\nonumber\\
& &\qquad + \int_0^{-\infty} d\psi\; \sinh\psi \exp\!\left(-2\pi i k_t s \cosh 
\psi\right)\bigg]\nonumber\\
& & + \pi \int_0^{-\infty} ds\;
s^2\; f\!\left(s^2\right) \bigg[\int_{\infty}^0
d\psi\; \sinh\psi \exp\!\left(-2\pi i k_t s \cosh \psi\right)\nonumber\\
& & \qquad +\int_{-\infty}^0 d\psi\; \sinh\psi \exp\!\left(-2\pi i k_t s \cosh 
\psi\right)\bigg]\nonumber\\ 
& = & \int_0^{\infty} ds\; s^2\; f\!\left( s^2\right)\; \frac{\exp\!\left(
-2\pi i k_t s\right)}{i k_t s} - \int^{-\infty}_0 ds\; s^2\; f\!\left( 
s^2\right)\; \frac{\exp\!\left(-2\pi i k_t s\right)}{i k_t s}\nonumber\\
& = & - \frac{2}{k_t} \int_0^{\infty} ds_0\; s_0\; f\!\left( s_0^2\right)\;
\sin\!\left( 2\pi k_t s_0\right),  
\eea
where
\beq
s_0 =\sqrt{ t^2-x^2-y^2}.
\eeq
The angular integrals have been computed as follows:
\bean
&&\int_0^{\infty} d\psi\; \sinh\psi \exp\!\left( - i a \cosh \psi\right)
+ \int_0^{-\infty} d\psi\; \sinh\psi \exp\!\left(-i a \cosh\psi\right)\\
&  &\qquad = \lim_{\epsilon\to 0} 2 \int_{0-i\epsilon}^{+\infty-i\epsilon}
d\psi\; \sinh\psi \exp\!\left(-i a \cosh\psi\right)\\
& & \qquad= \lim_{\epsilon\to 0} \frac{2}{i a}\bigg[ - \exp\!\left( -i a \cosh
\left(\infty-i\epsilon\right)\right) + \exp\!\left(-ia\cosh\left( 0-i\epsilon
\right)\right)\bigg]\\
& & \qquad=\frac{2}{i a} \exp\!\left( -i a\right),
\eean
where $a=2\pi k_t s$, and in the last line $\cosh(\psi\pm i\epsilon) = 
\cosh\psi \pm i \epsilon \sinh\psi$ has been used.

\item[(ii)] Spacelike momentum: $k_t=k_y=0$,
\bea
I_{I+III}(k_x) & = & \int_{I+III} dt\, dx\, dy\; f\!\left(t^2-x^2-y^2\right)\;
\exp\!\left( -2\pi i k_x x\right)\nonumber\\
& = & \int_0^{\infty} ds\; s^2\; f\!\left(s^2\right) 
\int_{-\pi/2}^{\pi/2} d\theta\; \bigg[\int_0^{\infty}
d\psi\; \sinh\psi \exp\!\left(-2\pi i k_x s \sinh \psi \cos\theta\right)
\nonumber\\
& &\qquad + \int_0^{-\infty} d\psi\; \sinh\psi \exp\!\left(-2\pi i k_x s \sinh 
\psi\cos\theta\right)\bigg]\nonumber\\
& & + \int_0^{-\infty} ds\;
s^2\; f\!\left(s^2\right) \int_{-\pi/2}^{\pi/2} d\theta 
\bigg[\int_{\infty}^0
d\psi\; \sinh\psi \exp\!\left(-2\pi i k_x s \sinh \psi\cos\theta\right)
\nonumber\\
& & \qquad +\int_{-\infty}^0 d\psi\; \sinh\psi \exp\!\left(-2\pi i k_x s \sinh 
\psi\cos\theta\right)\bigg]\nonumber\\ 
& = & \int_0^{\infty} ds\; s^2\; f\!\left( s^2\right)\; \frac{\exp\!\left(
-2\pi  k_x s\right)}{k_x s} + \int^{-\infty}_0 ds\; s^2\; f\!\left( 
s^2\right)\; \frac{\exp\!\left(-2\pi  k_x s\right)}{i k_x s}\nonumber\\
& = & \frac{2}{k_x} \int_0^{\infty} ds_0\; s_0\; f\!\left( s_0^2\right)\;
\exp\!\left(-2\pi k_x s_0\right).  
\eea
\end{itemize}
The angular integrals have been computed as follows ($a=2\pi k_x s$):
\bean
&&\int_{-\pi/2}^{\pi/2}d\theta\;\bigg[\int_0^{\infty} d\psi\; \sinh\psi\;
 \exp\!\left( -i a \sinh\psi \cos\theta\right)
  +  \int_0^{-\infty} d\psi\; \sinh\psi\; \exp\!\left(-i a\sinh \psi
\cos\theta\right)\bigg]\\
& &\qquad\qquad = 4 \int_0^{\infty} d\psi\;
\sinh\psi\; \int_0^{\pi/2} d\theta\; \cos\!\left( a\sinh\psi\cos\theta\right)\\
& & \qquad\qquad =  2\pi\int_0^{\infty}d\psi\; \sinh\psi\; J_0\left(a \sinh\psi\right)\\
& & \qquad\qquad = \frac{2\pi}{a} \exp(-a).
\eean
To get to the third line we have used formula 3.753(2) from Ref.~5, after
a change of variable to $x=\cos\theta$. The last line is 
obtained using 6.554(1) from Ref.~5, after a change of variable 
to $y=\sinh\psi$.

\item Patch II+IV:
\begin{itemize}
\item[(i)] Timelike momentum: $k_x=k_y=0$,
\bea
I_{II+IV} (k_t) & = & \int_{II+IV} dtdxdy\; f\!\left(t^2-x^2-y^2\right)\;
\exp\!\left(-2\pi i k_t t\right)\nonumber\\
& = & i\int_{i0}^{i\infty} ds\; s^2\; f\!\left(s^2\right)\; 
\int_{-\pi/2}^{\pi/2}d\theta\; \int_{-\infty}^{+\infty} d\psi\;
\cosh\psi\; \exp\!\left( 2\pi k_t s\sinh\psi\right)\nonumber\\
& &\qquad + i \int_{-i\infty}^{i0}ds\; s^2\; f\!\left( s^2\right)\;
 \int_{-\pi/2}^{\pi/2} d\theta\;
\int_{-\infty}^{+\infty} d\psi\; \cosh\psi\; \exp\!\left( 2\pi k_t s \sinh\psi
\right)\nonumber\\
& = & -\pi\int_{0}^{-\infty} ds'\; {s'}^2\; f\!\left(-{s'}^2\right)\; 
 \int_{-\infty}^{+\infty} d\psi\;
\cosh\psi\; \exp\!\left( -2\pi i k_t s'\sinh\psi\right)\nonumber\\
& &\qquad - \pi \int_{\infty}^{0}ds'\; {s'}^2\;  f\!\left(-{s'}^2\right)\;
\int_{-\infty}^{+\infty} d\psi\; \cosh\psi\; \exp\!\left( -2\pi i k_t s'
 \sinh\psi\right)\nonumber\\
& = & 0.
\eea
The angular integrals have been computed as follows (with $a=2\pi k_t s'$):
\bean
&&\int_{-\infty}^{+\infty} d\psi\; \cosh\psi \exp\!\left( - i a \sinh \psi\right)
\\
&  &\qquad = \lim_{\epsilon\to 0}  \int_{-\infty-i\epsilon}^{+\infty-i\epsilon}
d\psi\; \cosh\psi \exp\!\left(-i a \sinh\psi\right)\\
& & \qquad= \lim_{\epsilon\to 0} \frac{i}{a}\bigg[ \exp\!\left( -i a \sinh\psi
\right)\bigg]^{+\infty-i\epsilon}_{-\infty-i\epsilon}\\
& & \qquad = \lim_{\epsilon\to 0}\frac{i}{a} 
\bigg[\exp\!\left(- i a\infty\right)\; \exp\!\left( -\epsilon a \infty\right) -
\exp\!\left(i a\infty\right) \; \exp\!\left(-\epsilon a \infty\right)\bigg]\\
& & \qquad = 0.
\eean

\item[(ii)] Spacelike momentum: $k_t=0$,
\bea
I_{II+IV} & = & \int_{II+IV} dt dx dy\; f\!\left(t^2-x^2-y^2\right)\;
\exp\!\left(-2\pi i k_x x\right)\nonumber\\
& = & i \int_{i 0}^{i\infty} ds\; s^2\; f\!\left( s^2\right)\; 
\int_{-\pi/2}^{\pi/2} d\theta\; \int_{-\infty}^{+\infty} d\psi\; \cosh\psi\;
\exp\!\left( 2\pi k_x s \cosh\psi \cos\theta\right)\nonumber\\
& & \quad +\; i \int_{-i\infty}^{i0} ds\; s^2\; f\!\left( s^2\right)\; 
\int_{-\infty}^{+\infty} d\psi\; \cosh\psi\; \int_{-\pi/2}^{\pi/2}
d\theta\; \exp\!\left( 2\pi k_x s \cosh\psi \cos\theta\right)\nonumber\\
& = & 4 \int_0^{\infty} ds_1\; s_1^2\; f\!\left(-{s_1}^2\right)\;
\int_0^{\pi/2} d\theta \int_0^{\infty} d\psi\; \cosh\psi\;
\exp\!\left( 2\pi i k_x s_1 \cosh\psi \cos\theta\right)\nonumber\\
& & \quad + 4\int_0^{\infty} ds_1\; s_1^2\; f\!\left(s_1^2\right)\;
\int_0^{\pi/2} d\theta\; \int_0^{\infty} d\psi\; \cosh\psi\;
\exp\!\left(-2\pi i k_x s_1 \cosh\psi \cos\theta\right)\nonumber\\
& = & 8 \int_0^{\infty} ds_1\; s_1^2\; f\!\left( s_1^2\right)\; 
\int_0^{\pi/2} d\theta\; \int_0^{\infty} d\psi\; \cosh\psi\; 
\cos\!\left( 2\pi  k_x s_1\right)\nonumber\\
& = & \frac{2}{k_x} \int_0^{\infty} ds_1\; s_1\; f\!\left(s_1^2\right)\;
\cos\!\left( 2\pi k_x s_1\right),
\eea
where
\[
s_1 = \sqrt{x^2+y^2-t^2}.
\]
The angular integral has been computed as follows (with $a=2\pi k_x s_1$):
\bean
&&\int_0^{\infty} d\psi\; \cosh\psi\; \int_0^{\pi/2} d\theta\;
\cos\!\left( a \cosh\psi\; \cos\theta\right)\\
& & \qquad\qquad = \frac{\pi}{2}\;\int_0^{\infty} d\psi\; \cosh\psi\;
J_0\left( a \cosh\psi\right)\\
& & \qquad\qquad = \frac{\pi}{a}\; \cos (a),
\eean
where in the second line formula 3.715(19) from Ref.~5, and in the last line
formula 6.554(3) from Ref.~5 have been used, after a change of
variable to $x=\cosh\psi$.
\end{itemize}
\end{itemize}
In summary\\

\begin{center}
\fbox{\parbox{13.5cm}{\bean
I\left(k_t\right) & = & -\frac{2}{k_t} \int_0^{\infty}
 ds_0\; s_0\; f\!\left(
s_0^2\right)\; \sin\!\left( 2\pi k_t s_0\right)\\
I\left(k_x\right) & = & \frac{2}{k_x} \int_0^{\infty} ds_0\; s_0\; 
f\!\left( s_0^2\right)\; \exp\!\left( -2\pi k_x s_0\right) +\frac{2}{k_x}
\int_0^{\infty} ds_1\; s_1\; f\!\left( s_1^2\right)\; \cos\!\left( 2\pi k_x s_1
\right).
\eean}}
\end{center}

\section{Case $\Rset^{1,n}$}

In this section we follow Codelupi's derivation${}^3$ of
the $1+n$ dimensional case. The idea is to derive a recursion relation
between the Fourier transform in $n$ and $n+2$ spatial dimensions, and
then use the explicit expressions found before to construct the general
case.

\subsection{Recursion relation}

Define the radius in $n$ spatial dimensions as follows:
\[
r^2 = \sum_{i=1}^n x_i^2.
\]
Formally, the function $f\!\left(s\right) = f\!\left(\sqrt{t^2-r^2}\right)$
 always looks the same,
independent of the number of spatial dimensions. Suppose we have spaces
with spatial dimensions $n=1$ to $n=m$. For each of these spaces exists a
transform
\beq
F^{(n)}(k,k_0) = \int_0^{\infty}dr\; \chi_n(r,k)\; G(r,k_0),
\eeq
where (proof given in the appendix)
\beq
\chi_n(r,k) = 2\pi \frac{r^{n/2}}{k^{n/2-1}}\; J_{n/2-1} (2\pi r k)
\label{chieq}
\eeq
and 
\beq
G(r,k_0) = \int_{-\infty}^{+\infty} dt\; f\!\left(\sqrt{t^2-r^2}\right)\;
 \exp\!\left(-2\pi i k_0 t\right).
\eeq
But formally, $G(r,k_0)$ looks the same for all cases, for example,
\beq
G(r,k_0) = \int_0^{\infty} dk\; \chi_m(k,r)\; F^{(m)}(k,k_0),
\eeq
assuming that in the inverse Fourier transform, the angular contribution can
also be integrated out.\\
Substitute in the line before,
\bea
F^{(n)}(k,k_0) & = & \int_0^{\infty} dr\; \chi_n(r,k)\; \int_0^{\infty} du\;
\chi_m(u,r)\; F^{(m)}(u,k_0)\nonumber\\
& = & \int_0^{\infty}du\; F^{(m)}(u,k_0) \int_0^{\infty}dr\; \chi_n(r,k)\;
\chi_m(u,r).\label{fn}
\eea
We can explicitly evaluate the second integral using formula 6.575(1) 
from Red.~5. The result is
\beq
\int_0^{\infty}dr\; \chi_n(r,k)\;\chi_m(u,r) = \frac{2 \pi^h}{\Gamma(h)}
u \left(u^2-k^2\right)^{h-1} \Theta(u-k)
\eeq
where $h=(m-n)/2$, $\Gamma(x)$ Euler's Gamma function and $\Theta$ the
step function. Equation (\ref{fn}) now takes on the form
\beq
F^{(n)}(k,k_0) = \frac{2\pi^h}{\Gamma(h)} \int_k^{\infty} du\; 
F^{(n+2h)}(u,k_0)\; u\; \left(u^2-k^2\right)^{h-1}.
\eeq
Considering the special case $h=1$, i.e. $m=n+2$, and taking the derivative
with respect to $k$ of both sides of this equation leads to the recursion
formula
\beq
F^{(n+2)}(k,k_0) = -\frac{1}{2\pi k}\; \frac{\partial}{\partial k} 
F^{(n)}(k,k_0).\label{recurs}
\eeq
So the problem is solved, at least in principle, once we find the
explicit expressions for $n=1$ and $n=2$. But as we will show in the next
section (again following Codelupi${}^3$), the recursion relation above
 will also allow us to find explicit formulas for the case of general $n$.

\subsection{Explicit expressions for $\Rset^{1,n}$}

Let us define
\bean
l_0 & = & \sqrt{k_0^2-k^2},\\
l_1 & = & \sqrt{k^2-k_0^2}.
\eean
The recursion relation Eq.(\ref{recurs}) can be rewritten in terms of
 $l_0$ and $l_1$,
\bea
F^{(n+2)}(l_0) & = & -\frac{1}{2\pi k} \frac{\partial}{\partial k} 
F^{(n)}(l_0) = \frac{1}{2\pi l_0} \frac{d}{d l_0} F^{(n)}(l_0),\\
F^{(n+2)}(l_1) & = & -\frac{1}{2\pi k} \frac{\partial}{\partial k} 
F^{(n)}(l_1) = -\frac{1}{2\pi l_1} \frac{d}{d l_1} F^{(n)}(l_1).
\eea
To find  expressions for general $n$ consider the following two cases
\begin{itemize}
\item[(a)] n-even,
\bea
F^{(n)}(l_0) & = & \left(\frac{1}{2\pi l_0} \frac{d}{d l_0}
\right)^{\frac{n}{2}-1}
 F^{(2)}(l_0)\nonumber\\
& = & (-1)^{\frac{n}{2}} 2\pi \int_0^{\infty} ds_0\; f\!\left(s_0\right)\; 
\frac{s_0^{\frac{n+1}{2}}}{l_0^{\frac{n-1}{2}}}\; J_{\frac{n-1}{2}}\left(
2\pi s_0 l_0\right).
\eea
The last equation is proved be iteratively applying the derivative to
$F^{(2)}(l_0)$ (result of Section III.2) and using formula 8.472(2) from
Ref.~5. Similarly, we find
\bea
F^{(n)}(l_1) & = &  4 \int_0^{\infty} ds_0\; f\!\left(s_0\right)\;
 \frac{s_0^{\frac{n+1}{2}}}{l_1^{\frac{n-1}{2}}}\; K_{\frac{n-1}{2}}\left(
2\pi s_0 l_1\right)\nonumber\\
& & - 2\pi \int_0^{\infty} ds_1\; f\!\left(s_1\right)\; 
\frac{s_1^{\frac{n+1}{2}}}{l_1^{\frac{n-1}{2}}}\; N_{\frac{n-1}{2}}\left(
2\pi s_1 l_1\right),
\eea
where the formulae 8.472(2) and 8.486(13) from Ref.~5 have been used.
\item[(b)] n-odd,
\bea
F^{(n)}(l_0) & = & \left( \frac{1}{2\pi l_0}\right)^{\frac{n-1}{2}}
F^{(1)}(l_0)\nonumber\\
& = & (-1)^{\frac{n+1}{2}} 2\pi \int_0^{\infty} ds_0\; f\!\left(s_0\right)\; 
\frac{s_0^{\frac{n+1}{2}}}{l_0^{\frac{n-1}{2}}}\; N_{\frac{n-1}{2}}\left(
2\pi s_0 l_0\right)\nonumber\\
& & + (-1)^{\frac{n-1}{2}} 4 \int_0^{\infty} ds_1\; f\!\left(s_1\right)\;
 \frac{s_1^{\frac{n+1}{2}}}{l_0^{\frac{n-1}{2}}}\; K_{\frac{n-1}{2}}\left(
2\pi s_1 l_0\right)
\eea
and
\bea
F^{(n)}(l_1) 
 & = & 4 \int_0^{\infty} ds_0\; f\!\left(s_0\right)\; 
\frac{s_0^{\frac{n+1}{2}}}{l_1^{\frac{n-1}{2}}}\; K_{\frac{n-1}{2}}\left(
2\pi s_0 l_1\right)\nonumber\\
& & -2\pi \int_0^{\infty} ds_1\; f\!\left(s_1\right)\;
 \frac{s_1^{\frac{n+1}{2}}}{l_1^{\frac{n-1}{2}}}\; N_{\frac{n-1}{2}}\left(
2\pi s_1 l_1\right),
\eea
using the results of Section II.2, and, again, formulas 8.472(2) and
 8.486(13) from Ref.~5. 
\end{itemize}
For even $n$, the Fourier transform with a timelike momentum has
no contribution from the spacelike region of spacetime.\\
We can summarize both cases in the following formulas, now valid for
arbitrary $n$:\\
\begin{center}
\fbox{\parbox{12cm}{\bean
 F^{(n)}(l_0) & = &  - 2\pi \int_0^{\infty} ds_0\;
 f\!\left(s_0\right)\; 
\frac{s_0^{\frac{n+1}{2}}}{l_0^{\frac{n-1}{2}}}\; \bigg[ 
 N_{\frac{n-1}{2}}\left(2\pi s_0 l_0\right)\; \cos\!\left( \pi \frac{n-1}{2}
\right)\nonumber\\
& & \qquad +  J_{\frac{n-1}{2}}\left(2\pi s_0 l_0\right) \sin\!\left(
\pi \frac{n-1}{2}\right)\bigg]\nonumber\\
& & + 4   \int_0^{\infty} ds_1\; f\!\left(s_1\right)\;
 \frac{s_1^{\frac{n+1}{2}}}{l_0^{\frac{n-1}{2}}}\; K_{\frac{n-1}{2}}\left(
2\pi s_1 l_0\right) \cos\!\left( \pi \frac{n-1}{2}
\right) , \\
F^{(n)}(l_1) & = &  4 \int_0^{\infty} ds_0\; f\!\left(s_0\right)\; 
\frac{s_0^{\frac{n+1}{2}}}{l_1^{\frac{n-1}{2}}}\; K_{\frac{n-1}{2}}\left(
2\pi s_0 l_1\right)\nonumber\\
 & & -2\pi \int_0^{\infty} ds_1\; f\!\left(s_1\right)\;
 \frac{s_1^{\frac{n+1}{2}}}{l_1^{\frac{n-1}{2}}}\; N_{\frac{n-1}{2}}\left(
2\pi s_1 l_1\right).
\eean}}
\end{center}

\pagebreak

\renewcommand{\thesection}{Appendix : Derivation of Eq.(\ref{chieq})}
\renewcommand\theequation{A\arabic{equation}}
\setcounter{equation}{0}

\section{}

Equation (\ref{chieq}) can be derived in several different ways. For direct 
integration see e.g., Ref.~6, Chap.~4. We follow here a derivation
presented in Ref.~7.

We define the Fourier transform on $\Rset^n$ by
\beq
F\!\left(k_1,\ldots,k_n\right) = \int_{-\infty}^{+\infty}\ldots
\int_{-\infty}^{+\infty} dx_1\ldots dx_n\; f\!\left( x_1,\ldots ,x_2\right)\;
\exp\!\left( -2\pi i \left(k_1 x_1 +\ldots +k_n x_n\right)\right),
\eeq
or short
\beq
F\!\left( \vec{k}\right) = \int_{\Rset^n} d^nx\; f\!\left(\vec{r}\right)\;
\exp\!\left(-2\pi i \vec{k}\cdot\vec{r}\right)
\eeq
and the inverse Fourier transform by
\beq
f\!\left( \vec{r}\right) = \int_{\Rset^n} d^nk\; F\!\left(\vec{k}\right)\;
\exp\!\left(2\pi i \vec{k}\cdot\vec{r}\right).
\eeq
One property we will need is
\beq
\Delta^2 f\!\left(\vec{r}\right) = \sum_m \frac{\partial^2 f}{\partial x_m^2}
= - 4 \pi^2 \int_{\Rset^n} d^nk\; k^2\; F\!\left(\vec{k}\right)\; 
\exp\!\left(2\pi i \vec{k}\cdot\vec{r}\right).\label{deltaf1}
\eeq

Now assume we have  a radial function, and we work in spherical coordinates.
We would like to write the Fourier transformation as
\beq
F(k) = \int_0^{\infty} dr\; f(r)\; \chi_n (r,k)
\eeq
and the inverse Fourier transform as 
\beq
f(r) = \int_0^{\infty} dk\; F(k)\; \chi_n (k,r) dk,\label{finvr}
\eeq
where the $\chi$'s contain the integration over the compact $n-1$ angular
coordinates.
From Eq.(\ref{deltaf1}) we also know
\beq
\Delta^2 f(r) = - 4 \pi^2 \int_0^{\infty} dk\; k^2\; \chi_n(k,r)\; F(k).
\label{deltaf2}
\eeq
Calculating $\Delta^2 f(r)$ (now starting with Eq.(\ref{finvr})),
\[
\frac{\partial^2 f(r)}{\partial x_m^2} = \int_0^{\infty} dk\; F(k)\;
\frac{\partial^2}{\partial x_m^2}\; \chi_n(k,r),
\]
where
\[
\frac{\partial^2}{\partial x_m^2} \chi_n (k,r) = \left(\frac{1}{r}
-\frac{x_m^2}{r^3}\right)\; \frac{\partial}{\partial r}\; \chi_n(k,r)
+ \frac{x_m^2}{r^2}\; \frac{\partial^2}{\partial r^2}\; \chi_n(k,r),
\]
yields
\beq
\Delta^2 f(r) = \sum_m \frac{\partial^2 f(r)}{\partial x_m^2} =
\int_0^{\infty} dk\; F(k)\; \left( \frac{n-1}{r}\;\frac{\partial}{\partial r}\;
\chi_n(k,r)+\frac{\partial^2}{\partial r^2}\; \chi_n\right).
\eeq
But with Eq.(\ref{deltaf2}),
\beq
\int_0^{\infty} dk\; F(k)\; \left[ \frac{\partial^2}{\partial r^2}\; 
\chi_n(k,r)+ \frac{n-1}{r}\; \frac{\partial}{\partial r}\; \chi_n(k,r)
+4\pi^2 k^2 \chi_n(k,r)\right] =0.
\eeq
This equation is valid for arbitrary $F(k)$, so the expression in 
brackets has to be zero. This ODE (in $r$) has the general solution
(see Ref.~8, p.146)
\beq
\chi_n(k,r) = A_n(k) r^{1-n/2} Z_p\left(2\pi k r\right)\label{chisol},
\eeq
where $A_n(k)$ is determined by the initial conditions, $Z$ is a Bessel
function of order $p$, and $p=\pm\left(1-n/2\right)$.
Computing the inverse Fourier transform explicitly in the cases of
$n=1$ and $n=2$, determines $p$ to be $n/2-1$.\\
To find $A_n$, consider $f(r)$ at $r=0$,
\[
f(0) = \int_0^{\infty} dk\; F(k)\; \chi_n(k,0).
\]
From Eq.(\ref{chisol}) we have
\bean
\chi_n(k,0) & = & \lim_{r\to 0} A_n(k) r^{1-n/2} J_{n/2-1}\left(2\pi r
 k\right)\\
& = & A_n(k)\; \frac{r^{1-n/2} (\pi r k)^{n/2-1}}{\Gamma(n/2)}\\
& = & A_n(k)\; \frac{(\pi k)^{n/2-1}}{\Gamma(n/2)},
\eean
where 9.1.7 from Ref.~9 has been used.\\
But according to the definition of the inverse Fourier transform
\bean
f(0) & = & \int_{\Rset^n} dk^n\; F(k)\\
& = & \frac{\pi^{n/2} n}{\Gamma\left(1+n/2\right)}\; \int_0^{\infty}
dk\; k^{n-1}\; F(k).
\eean
The factor in front of the integral is the volume of the unit $n-1$-sphere.
Equating both expressions for $f(0)$, which are valid for arbitrary $F(k)$,
yields
\[
A_n(k) = 2\pi k^{n/2},
\]
and therefore as the final result
\beq
\chi_n(k,r) = 2\pi k^{n/2} r^{1-n/2} J_{n/2-1}\left(2\pi r k\right).
\eeq
Because of the symmetry of the transformation,
\beq
\chi_n(r,k) = 2\pi r^{n/2} k^{1-n/2} J_{n/2-1}\left(2\pi r k\right).
\eeq

\pagebreak

\newcounter{footn}
\begin{list}{${}^{\arabic{footn}}$}{\usecounter{footn} 
\setlength{\topsep}{1.5cm}}
\item
P.~Cartier and C.~DeWitt-Morette, ``A new perspective on functional
integration,'' J.~Math.~Phys. {\bf 5}, 2237 (1995).
\item
M.E.~Peskin and D.V.~Schroeder, {\it An Introduction to Quantum Field
Theory} (Addison-Wesley, New York, 1995).
\item
R.~Codelupi, ``Transformate di Fourier di funzioni invarianti rispetto
al gruppo di Lorentz,'' Note Recensioni e Notizie {\bf 18}, 853
(1969).
\item
N.~Krausz and M.S.~Marinov, ``Exact evolution operator on noncompact
group manifolds,'' J.~Math.~Phys. {\bf 41}, 5180 (2000).
\item
I.S.~Gradshteyn and I.M.~Ryzhik, {\it Table of Integrals, Series, and
 Products} (Corrected and enlarged edition) (Academic, New
York, 1980).
\item
E.M.~Stein and G.~Weiss, {\it Introduction to Fourier Analysis on Euclidean
Spaces} (Princeton University Press, Princeton, NJ, 1971).
\item
R.~Codelupi, ``Transformate di Fourier pluridimensionali,''
 Note Recensioni e Notizie {\bf 18}, 311 (1969).
\item
E.~Jahnke and F.~Emde, {\it Tables of Functions with Formulae and Curves}
 (4th edition) (Dover, New York, 1945).
\item
M.~Abramowitz and I.A.~Stegun, {\it Handbook of Mathematical Functions}
(Dover, New York, 1965).
\end{list}

\end{document}